\documentclass[aps,pra,twocolumn,showpacs,amsmath,amssymb,preprintnumbers,superscriptaddress,10pt,longbibliography]{revtex4-1}

\usepackage{bm}
\usepackage{graphicx}
\graphicspath{{figures/}}
\usepackage{SIunits}
\usepackage{hyphenat}
\usepackage{physics}
\usepackage{scrextend}
\usepackage[bookmarks=false]{hyperref}
\usepackage{color}
\usepackage[capitalise]{cleveref}
\crefname{section}{Sec.}{Secs.}% APS style uses abbreviations
\Crefname{section}{Section}{Sections}
\newcommand{\tran}{^\top}

\definecolor{pink}{RGB}{255,0,255}

\definecolor{blue}{rgb}{0,0,1}

\definecolor{red}{rgb}{1,0,0}

\begin{document}

\title{Characterisation of state-preparation uncertainty in quantum key distribution}

\author{Anqi~Huang}
\email{angelhuang.hn@gmail.com}
\affiliation{Institute for Quantum Information \& State Key Laboratory of High Performance Computing, College of Computer Science and Technology, National University of Defense Technology, Changsha 410073, People's Republic of China} % Country name is spelled this way per APS style

\author{Akihiro~Mizutani}
\affiliation{Graduate School of Engineering Science, Osaka University, Toyonaka, Osaka 560-8531, Japan}

\author{Hoi-Kwong~Lo}
\affiliation{\mbox{Centre for Quantum Information and Quantum Control (CQIQC), Department of Electrical \& Computer Engineering} \mbox{and Department of Physics, University of Toronto, Toronto, Ontario M5S~3G4, Canada}}
\affiliation{Department of Physics, University of Hong Kong, Pokfulam, Hong Kong}
\affiliation{Quantum Bridge Technologies, Inc.,\ 100 College Street, Toronto, Ontario M5G~1L5, Canada}

\author{Vadim~Makarov}
\affiliation{Russian Quantum Center, Skolkovo, Moscow 121205, Russia}
\affiliation{\mbox{Shanghai Branch, National Laboratory for Physical Sciences at Microscale and CAS Center for Excellence in} \mbox{Quantum Information, University of Science and Technology of China, Shanghai 201315, People's Republic of China}} % Country name is spelled this way per APS style
\affiliation{NTI Center for Quantum Communications, National University of Science and Technology MISiS, Moscow 119049, Russia}

\author{Kiyoshi~Tamaki}
\affiliation{Faculty of Engineering, University of Toyama, Gofuku 3190, Toyama 930-8555, Japan}

\date{\today}

\begin{abstract}
%If the journal requires a shorter abstract under 600 characters: To achieve secure quantum key distribution, all imperfections in the source unit must be incorporated in a security proof and measured in the lab. Here we perform a proof-of-principle demonstration of the experimental techniques for characterising the source phase and intensity fluctuation in commercial quantum key distribution systems. We then apply the experimental results to the security proof that takes into account fluctuations in the state preparation and study the resulting secure key rates. Our characterisation methods pave the way for a future certification standard.
To achieve secure quantum key distribution, all imperfections in the source unit must be incorporated in a security proof and measured in the lab. Here we perform a proof-of-principle demonstration of the experimental techniques for characterising the source phase and intensity fluctuation in commercial quantum key distribution systems. When we apply the measured phase fluctuation intervals to the security proof that takes into account fluctuations in the state preparation, it predicts a key distribution distance of over $100~\kilo\meter$ of fiber. The measured intensity fluctuation intervals are however so large that the proof predicts zero key, indicating a source improvement may be needed. Our characterisation methods pave the way for a future certification standard.
\end{abstract}

\maketitle

\section{Introduction}
\label{sec:intro}

Quantum key distribution~(QKD), which guarantees information-theoretically secure communication in theory, has become one of the most widely applied techniques of quantum information processing~\cite{gisin2002, lo2014}. However, practical imperfections in devices challenge the theoretically-guaranteed security during QKD global-scale deployment~\cite{xu2010,lydersen2010a,sun2011,jain2011,huang2016,huang2019,huang2020,wu2020,sun2022,gao2022}. Fortunately, a measurement-device-independent~(MDI) QKD protocol~\cite{lo2012} is immune to all measurement-side secure loopholes, because this protocol does not make assumption on the measurement devices. Thus, the loopholes in the source unit are the last obstacle to achieve security of QKD in reality~\cite{jain2014,huang2018,huang2019,huang2020,ponosova2022}. The most effective solution to eliminate such security threat is to consider the practical imperfections of the source in the security model.

The most widely used ideal source model is the one that emits only a single photon without encoding errors, and based on this model, numerous security proofs have been made~\cite{shor2000,tomamichel2012,tomamichel2017}. Unfortunately, however, this model does not properly reflect the actual properties of practical sources. One imperfection is that a practical source sometimes emits multiple photons. Another is that an encoding operation inevitably entails modulation inaccuracies owing to limited precision of experimental devices. These imperfections must be incorporated in the security proofs. Thanks to the invention of the decoy-state method \cite{wang2005a,ma2005,lo2005}, where Alice probabilistically varies the mean photon number of phase-randomised coherent light and it only matters that the source has a sufficiently large portion of single-photon emission, we can remove the requirement of the single-photon source \footnote{Note that a basic assumption in the decoy-state method is that signal- and decoy-pulses are indistinguishable, but this assumption has been waived in Refs.~\onlinecite{huang2020,mizutani2019}.}. To tackle the encoding inaccuracies, two approaches have been proposed. The first one is to take this into account by considering the fidelity between the $Z$- and $X$-basis states \cite{gottesman2004,lo2007,koashi2009}. The second one is to employ a loss-tolerant protocol where basis-mismatched events are exploited besides the basis-matched ones for a better estimation of leaked information \cite{tamaki2014,boaron2018,grunenfelder2018}. The loss-tolerant protocol has a remarkable property that the secret key is tolerant to the channel loss even if the source emits a state deviated from the expected one. Owing to this advantage, this protocol has attracted intensive attention in theory \cite{xu2015,mizutani2015,nagamatsu2016,mizutani2019} and experiment \cite{xu2015,tang2016a}. The original loss-tolerant protocol \cite{tamaki2014} has been made more practical by generalising it to the case where Alice knows only the intervals of the phase and intensity fluctuations of the coherent source \cite{nagamatsu2016,mizutani2019}. Long-distance QKD using the loss-tolerant protocol has been realised under realistic intervals of phase and intensity fluctuations \cite{xu2015,mizutani2015,mizutani2019,lu2019,lu2021}.

To apply these security proofs to calculate the key rate of a running QKD system, one shall obtain some parameters from the QKD system during raw key exchange as the inputs of the calculation. For instance, the detection gain and qubit error rate~(QBER) should be observed in order to calculate the secure key rate in the decoy-state protocol~\cite{ma2005}. Similarly, the trace distance should be known before quantifying the amount of leaked information from the source~\cite{tamaki2016, huang2018}. For the imperfect state encoding, the fluctuation or deviation of state modulation should be characterised as a crucial value for key-rate calculation. Reference~\onlinecite{xu2015} experimentally measures the averaged phase deviation of modulation and applies it to the loss-tolerant protocol~\cite{tamaki2014} with finite-size analysis. However, there is no methodology so far to characterise the fluctuation interval of the imperfect modulation in a practical QKD system~\cite{mizutani2019}.

Here we propose two methods of experimentally characterising fluctuation, one for phase and another for intensity. We apply each of them to a QKD system to obtain the intervals of phase and intensity fluctuation. By following the loss-tolerant protocol in Ref.~\onlinecite{mizutani2019}, we treat the optical pulses that lie in this interval as untagged signals and the others as tagged signals. Note that the secret key is extracted from the untagged signals, and the information of the tagged signals is completely leaked to an eavesdropper. The simulation shows that secret key can be produced between Alice and Bob over more than $100~\kilo\meter$ of fiber with $10^{13}$ pulses sent, when the phase fluctuation only is considered. However the measured intensity fluctuation is too large for the secret key to be produced, owing to the relatively unstable type of source used in the QKD system we have tested (a gain-switched semiconductor laser). We therefore simulate performance of a system that has the actual measured phase fluctuation and a fraction of the measured intensity fluctuation, to demonstrate how much the source has to be improved.

The rest of this article is organised as follows. In \Cref{sec:assumptions} we state all the assumptions on the QKD system in our method. \Cref{sec:experiment-phase} presents the methodology of characterising the phase fluctuation on an experimental example of a modified Clavis2 QKD system from ID~Quantique. \Cref{sec:experiment-intensity} presents the methodology of characterising the intensity fluctuation on a different prototype QKD system. The experimentally obtained values of phase and intensity fluctuation are applied to the security proof in \cref{sec:simulation}. We conclude in \cref{sec:conclusion}.

\section{Assumptions in experiments}
\label{sec:assumptions}

To characterise the uncertainty of state preparation, the intervals of phase fluctuation are measured on a commercial plug-and-play phase-encoding QKD system Clavis2 from ID Quantique \cite{idqclavis2specs}. Since no decoy states are employed in the Clavis2 system, the intervals of intensity fluctuation are measured on another prototype QKD system running a decoy-state BB84 protocol with polarization encoding. The phase and intensity intervals are measured on two separate QKD systems because we have had no access to a QKD system that employs a phase-encoding loss-tolerant protocol with decoy-state method. However, the methodology of characterisation proposed in this work is general and applicable to the loss-tolerant QKD systems.

The phase and intensity fluctuations are considered in the security proof proposed in Ref.~\onlinecite{mizutani2019}. For those not familiar with the security model in Ref.~\onlinecite{mizutani2019}, the framework is recapped in~\cref{sec:protocol}. In this security proof, the source's phase and intensity fluctuation falls within a phase interval $R_\text{ph}^{c}$ and intensity interval $R^k_{\rm int}$ with probability at least $\left(1 - p_\text{fail}\right)$. A value outside either $R_\text{ph}^{c}$ or $R^k_{\rm int}$ is regarded as an error, which happens with probability at most $p_\text{fail}$. Given a mean value $m$ of the phase or intensity for a certain state the source prepares, the interval may be defined as $\left(1 \pm x\%/100\%\right) m$, or briefly $\pm x\%$, where $x$ is an arbitrary factor. If a statistical distribution of the emitted states is measured and found to be Gaussian, as is the case in our experiments, the interval may alternatively be defined as $m \pm y\sigma$, or briefly $\pm y\sigma$, where $\sigma$ is the standard deviation and $y$ an arbitrary factor. The error probability $p_\text{fail}$ has to be a small value, usually on the order of $10^{-9}$. These are the only two parameters required by the proof. 

In order to ensure our characterisation is valid, below we summarise all the assumptions we make in the experiments.

(1)~The phase and intensity values obtained by measuring bright pulses with classical photodetectors are identical to these of pulses attenuated to a single-photon level.

(2)~The laser generates single-mode pulses with the randomised common phase of signal and reference pulses, which corresponds to Assumption~(A-1) in~\cref{sec:protocol}.

(3)~The phase distribution for each choice of $c^i \in \mathcal{C} := \{0_Z, 1_Z, 0_X\}$ and the intensity distribution for each choice of $k^i \in \mathcal{K} := \{k_1, k_2, k_3\}$ are identically and independently distributed, although our security model can handle the setting-choice-independent correlation~(SCIC) that is explained in (A-2) in~\cref{sec:protocol}.

(4)~The classical photodetectors are well characterized to linearly convert optical power to electrical voltage, and the oscilloscopes measure the input electrical voltage linearly.

(5)~In our experiment, we first measure the phase value in each of about $10^5$ samples (or intensity value in $1.5 \times 10^5$ signal and $2.5 \times 10^4$ decoy and vacuum states each). Then we plot the resulting phase (or intensity) distributions. For simplicity, we assume that these sample sizes are large enough to ignore statistic noise.

\begin{figure*}
\includegraphics{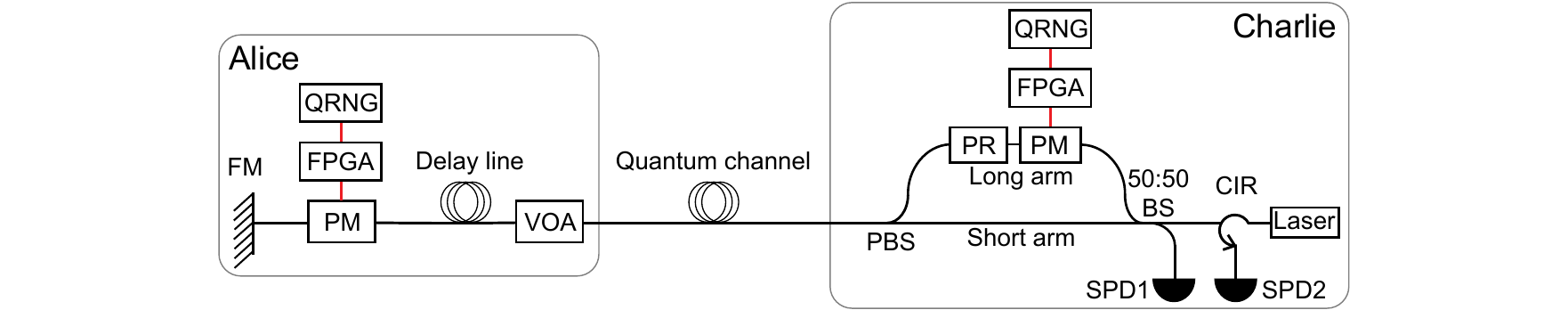}
\caption{\label{fig:setup-Clavis2}Scheme of Clavis2 plug-and-play QKD system. Alice's PM controlled by a field-programmable gate array~(FPGA) randomly applies a phase $\theta_\text{A} \in \{0, \pi/2, \pi, 3\pi/2 \}$, and Charlie's PM controlled by another FPGA randomly applies a phase $\theta_\text{B} \in \{0, \pi/2 \}$. The quantum random number generators~(QRNGs) provide random numbers to the FPGAs, in order to randomly select the phases. The $90\degree$ polarization rotator (PR) in Charlie is not a discrete component but is implemented by effectively `twisting' the fiber in the plane of the drawing, such that the light polarizations in the two arms become orthogonal at the PBS. In practice, this is achieved by aligning the axes of the polarization-maintaining fiber pigtails appropriately between the PBS and PM. SPD: single-photon detector; BS: beam splitter; PBS: polarization beam splitter; FM: Faraday mirror; CIR: circulator; VOA: variable optical attenuator.}
\end{figure*}

\begin{figure*}
\includegraphics{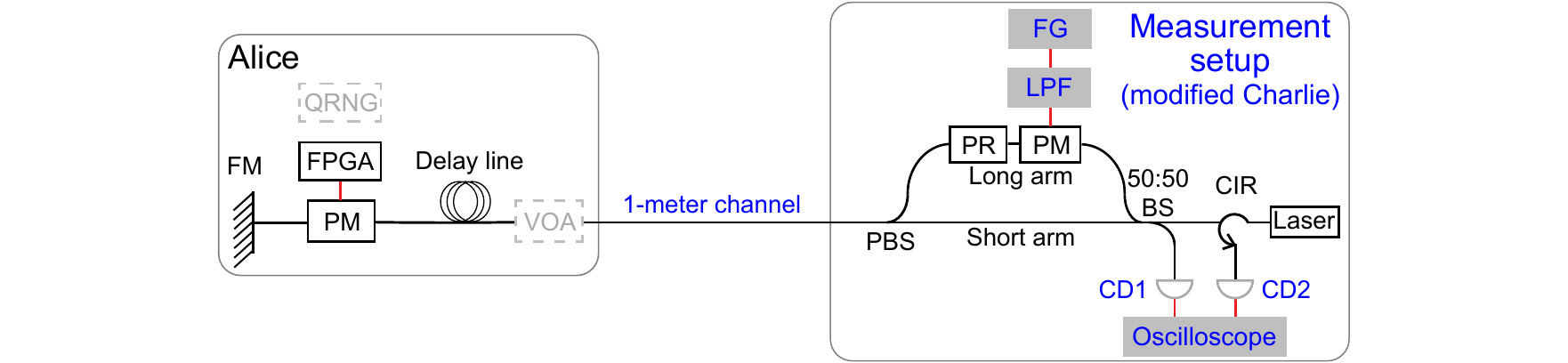}
\caption{\label{fig:setup-phase}Setup for phase distribution measurement. Alice and Charlie are connected by a 1-meter fiber. The device under test, Alice's phase modulator (PM), is still controlled by the FPGA in the system, and Charlie's PM is controlled by an external function generator (FG). The replacements required for this measurement are highlighted in grey; the VOA and the QRNG in Alice are inactivated. See text for details. CD: classical photodetector (optical-to-electrical converter); LPF: lowpass filter.}
\end{figure*}

(6)~The phase $\theta_\text{A} = 0$ is assumed to be a reference state, relative to which the other states are encoded. We then characterise the noise at $\theta_\text{A} = 0$ and remove it from the other states. This removes both fluctuation introduced by the remaining parts of the experiment and any possible fluctuation of the reference state itself. Regarding intensity modulator~(IM), it is assumed to be free of fluctuation when the QKD system is powered on but does not run a raw key exchange. The measured results for the case of no key exchange then characterise the fluctuation from the remaining parts of the experiment. Also, we assume that these fluctuations are the same during our measurements of the other values of $\theta_\text{A}$ or intensity. Therefore, we can use a signal processing technique to remove the noise.

(7)~The filtering technique of singular-value-decomposition (SVD; see \cref{sec:SVD}) is assumed to work sufficiently well to remove most of the noise and output distributions of phase and intensity that are close to the real ones.

(8)~The fluctuation measured from the nonrandom phase modulation is assumed to be identical to that of random phase modulation during the stage of raw key exchange in QKD protocol.

\section{Experiment of phase characterisation}
\label{sec:experiment-phase}

In this work, we first focus on the experimental characterisation of phase fluctuation. The characterised phase interval $R_\text{ph}^{c}$ is applied to estimate the upper bound on the number of phase errors per Eq.~(24) in Ref.~\onlinecite{mizutani2019}. In order to study the methodology of fluctuation measurement and have a rough idea about the values of phase fluctuation interval $R_\text{ph}^{c}$ in practical QKD systems, we conduct a proof-of-principle measurement to characterise this parameter. The phase interval $R_\text{ph}^{c}$ is measured in the commercial QKD system Clavis2 from ID~Quantique~\cite{idqclavis2specs}, which is a plug-and-play scheme~\cite{muller1997, stucki2002}. Importantly, this scheme is quite suitable to measure the phase value, because it is inherently stable without active calibration---the phase drift, polarization birefringence, and laser intensity fluctuations are automatically compensated. 

The Clavis2 system works as follows~(see~\cref{fig:setup-Clavis2}). To avoid confusion, we call Bob in this system as Charlie hereafter, who performs the characterisation of phase fluctuation instead of acting as the receiver of a running QKD system like Bob. The laser emits linear-polarization pulses in a single mode with randomised phases at the repetition rate of $5~\mega\hertz$. Each pulse then splits to be two pulses~(signal pulse in the long arm and reference pulse in the short arm with $50~\nano\second$ timing difference), combined and sent to Alice via an optical fiber. At this point, the signal and reference pulses are in the vertical and the horizontal polarizations, respectively. Alice modulates only the phase of the signal pulse, reflects both pulses with a Faraday mirror~(FM), and sends them back to Charlie after attenuating the pulses to single-photon level. Charlie modulates the phase of the reference pulse (that is now routed into the long arm), and then the interfered signals are detected by two single-photon detectors~(SPDs). The delay line in Alice is used to store a packet of 1700 incoming pulse pairs~(called a frame), thus avoiding an intersection between the incoming and reflected pulses in the quantum channel.

In the experiment, we would like to characterise the phase fluctuation $R_\text{ph}^{c}$ of the built-in PM in Alice in the Clavis2 system. Note that the previous work in Ref.~\onlinecite{xu2015} measured only the mean value of the modulated phase, but not its distribution. In order to measure the latter, we modify Bob in Clavis2 to be a Charlie module as shown in \cref{fig:setup-phase}, which acts as a characterisation apparatus. We design this measurement apparatus accordingly to apply a stable, calibrated, passively low-pass-filtered phase shift to the entire frame of the incoming light pulses to do a correct measurement. In this figure, replacements are highlighted in grey, and inactivated components are in grey dashed boxes. In order to obtain pulses with high intensities, we set Alice's VOA to its lowest possible setting~(about $2~\deci\bel$). The SPDs at the outputs of Charlie's interferometer are replaced with two classical optical-to-electrical~(O/E) converters (LeCroy OE455 with $3.5$-$\giga\hertz$ electrical bandwidth) connected to an oscilloscope (LeCroy 760Zi with $6$-$\giga\hertz$ analog bandwidth and $20$-$\giga\hertz$ sampling rate). In order to separately measure the distribution of each phase value, we inactivate the QRNG in Alice and apply the same phase value to every pulse. Charlie's PM is controlled by an external function generator~(FG; Highland Technology P400) to provide a stable, fluctuation-free phase $\theta_\text{B} \in \{0, \pi/2\}$. We will explain this point in more detail later. A passive lowpass filter~(LPF) with $20~\kilo\hertz$ cutoff is inserted between the output of the function generator and Charlie's PM, ensuring a low-noise electrical signal at the latter. Note that in the plug-and-play system used here, one cannot simply apply a static phase shift in the modified Charlie, as it would have canceled out when the frame of light pulses goes out of Charlie then returns. However, a single long phase-shifting electrical pulse externally applied at the modified Charlie's PM well in advance of the returning frame is sufficient to perform an accurate measurement in this system configuration.

The procedure of the experiment is as follows.

\medskip\noindent\emph{Stage 0.\ Preset.}\nopagebreak

This stage mainly focuses on preparing the electronic signal applied to Charlie's PM to serve a stable modulation, which ensures the fluctuation we later measure in the experiment is only from Alice's PM, but not from that of Charlie. When a frame of pulses comes back to Charlie, the function generator applies a $500$-$\micro\second$ long voltage pulse that covers the entire $350$-$\micro\second$ frame, avoiding frequent rising and falling edges and thus avoiding oscillations in the voltage. Also, this voltage pulse is applied slightly before the frame and ends after the frame. Thus, no optical pulse is modulated during the rising and falling edges of the voltage pulse, and an identical phase is applied to all the optical pulses in the frame. Furthermore, the $20~\kilo\hertz$-bandwidth LPF mitigates small fluctuation due to any electronic noise, resulting in a steady voltage level. This strategy of providing a stable modulation to Charlie's PM is applied to both the following calibration and measurement. However, we cannot adopt this technique to Alice's PM because the phase is not randomly modulated pulse-by-pulse, which does not satisfy the fundamental assumption of QKD.

\medskip\noindent\emph{Stage 1.\ Calibration.}\nopagebreak

Calibration is necessary to perform precise measurement and determine the voltage value applied to Charlie's PM for $\theta_\text{B} = \pi/2$. First, we set both $\theta_\text{A} = 0$ and $\theta_\text{B} = 0$. Ideally, this would result in a full interference: CD2 receives all the energy, but zero energy measured by CD1. However, in practice, the energy detected by CD1 is not zero, due to imperfect alignment between Alice and Charlie. Thus, we denote the output energy at CD1 under this misalignment by $D_{1,\text{min}}$, meanwhile the output energy measured by CD2 is maximum, denoted by $D_{2,\text{max}}$. Then we keep $\theta_\text{A} = 0$ and scan the voltage applied to Charlie's PM. During the scanning, the maximum output energy measured by CD1 is denoted by $D_{1,\text{max}}$, and at this moment the output energy measured by CD2 is minimum, denoted by $D_{2,\text{min}}$. It is notable that $D_{2,\text{max}}$ is slightly smaller than $D_{1,\text{max}}$ because the circulator between the interferometer's output and CD2 introduces loss. A ratio $R=D_{1,\text{max}}/D_{2,\text{max}}$ quantifies this extra loss, which we compensate in the phase calculation in Stage~3.

To obtain a precise voltage for $\theta_\text{B} = \pi/2$, we gradually increase the voltage applied to Charlie's PM from zero~(while keeping $\theta_\text{A} = 0$), until we detect the energy of $(D_{1,\text{max}} - D_{1,\text{min}})/2$ at CD1 and the energy of $(D_{2,\text{max}} - D_{2,\text{min}})/2$ at CD2. These happen simultaneously. This moment implies the relative phase between Alice and Charlie is $\pi/2$. Since we have set $\theta_\text{A} = 0$, we can deduce that $\theta_\text{B} = \pi/2$.

\begin{figure}
\includegraphics{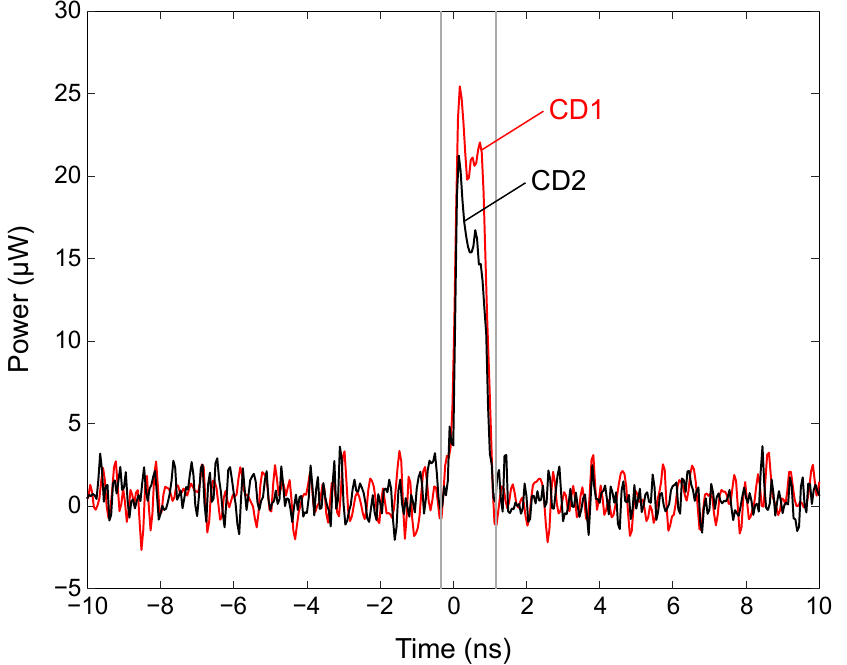}
\caption{\label{fig:waveform-phase-pi-2}Typical waveforms measured by CD1 and CD2 for $\theta_\text{A}=\pi/2$. The energy of each pulse is calculated by integrating the power between the grey lines.}
\end{figure} 

After this calibration, we perform the following stages for each phase value $\theta_\text{A} \in \{0, \pi/2, \pi \}$.

\medskip\noindent\emph{Stage 2.\ Measurement.}\nopagebreak

In order to measure the fluctuation of each $\theta_\text{A}$, Alice's PM modulates every pulse individually but applies the same phase. Charlie's PM applies either $\theta_\text{B} = 0$ (for Alice's phase $\theta_\text{A}=\pi/2$) or $\theta_\text{B} = \pi/2$ (for $\theta_\text{A}=0$ or $\pi$) to ensure that phase difference between Alice and Charlie is always $\pi/2$ no matter which $\theta_\text{A}$ Alice selects. The $\pi/2$ working point is necessary for the measurement, because at the 0 and $\pi$ points of full interference, the derivative of the output energy would tend to zero and not allow phase extraction. On the other hand, the point at $\pi/2$ can provide the maximal derivative of the output energy as a function of the phase, allowing us to perform an accurate measurement. Typical waveforms measured by CD1 and CD2 are shown in~\cref{fig:waveform-phase-pi-2}. The measured energy of each pulse is calculated by integrating the power between the grey lines. The whole system keeps running until we collect $10^5$ pulses for each phase value.

\begin{figure*}
\includegraphics{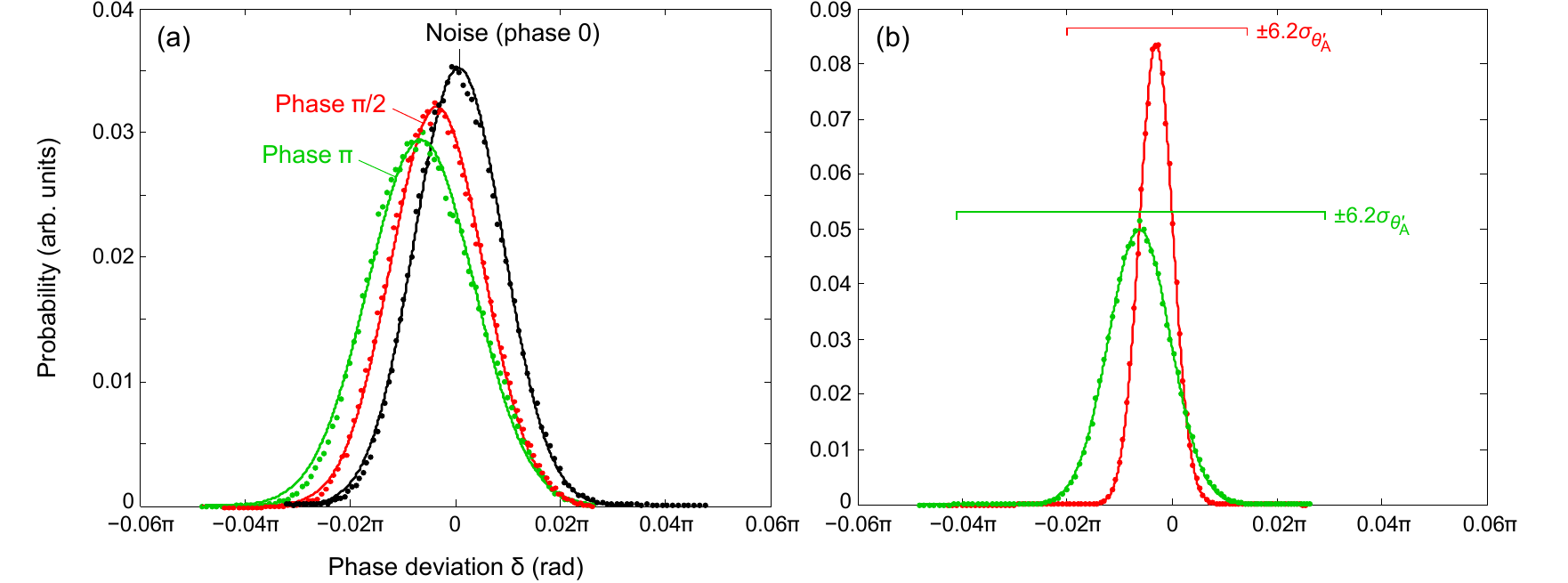}
\caption{\label{fig:distribution-phase}Empirical distribution of phase deviation $\delta$ for different phases Alice applies, (a) before and (b) after noise filtering at Stage~4. The dots are measured data, and the curves are Gaussian approximations. Intervals $R_\text{ph}^{c}$ are shown as bars.}
\end{figure*}

\medskip\noindent\emph{Stage 3.\ Phase calculation.}\nopagebreak

Let us denote the energy measured by the two detectors as $D_{1,\theta_\text{A}}$ and $D_{2,\theta_\text{A}}$. The value of the actual phase, $\theta'_\text{A}$, can be calculated after we compensate the extra loss for CD2 by multiplying its energy by the ratio $R$ obtained at Stage~1 and subtract the global misalignment between Alice and Charlie. Thus, the phase deviation $\delta$ between the real phase $\theta'_\text{A}$ and the ideal phase $\theta_\text{A}$ can be calculated as~\footnote{The outputs of the Mach-Zehnder interferometer are
$I_1 = I\sin^2 \left((\theta_A'-\theta_B)/2\right)$,
$I_2 = I\cos^2 \left((\theta_A'-\theta_B)/2\right)$,
where $I$ is the initial intensity of the source. Therefore, the ratio $I_1/I_2 = \tan^2\left((\theta_A'-\theta_B)/2\right)$. Since we compensate the extra loss and subtract the misalignment in our experiment, $I_1= D_{1,\theta_\text{A}}-D_{1,\text{min}}$ and $I_2 = R\left(D_{2,\theta_\text{A}}-D_{2,\text{min}}\right)$. Therefore, we can obtain the phase deviation as shown in~\cref{deviation}.}
\begin{widetext}
\begin{equation}
\label{deviation}
\delta = \theta'_\text{A} - \theta_\text{A} =\begin{cases}
2\arctan\left(-\sqrt{\frac{D_{1,\theta_\text{A}}-D_{1,\text{min}}}{R\left(D_{2,\theta_\text{A}}-D_{2,\text{min}}\right)}}\right) -\Delta, \text{~if $\theta'_\text{A} - \theta_\text{B} < 0$}\\
~\\
2\arctan\sqrt{\frac{D_{1,\theta_\text{A}}-D_{1,\text{min}}}{R\left(D_{2,\theta_\text{A}}-D_{2,\text{min}}\right)}} -\Delta, \text{~if $\theta'_\text{A} - \theta_\text{B} > 0$}
\end{cases},
\end{equation}
\end{widetext}
where $\Delta=\theta_\text{A}-\theta_\text{B}$ is the nominal phase difference between Alice and Charlie. According to the measured $D_{1,\theta_\text{A}}$ and $D_{2,\theta_\text{A}}$ over $10^5$ pulses for each $\theta_\text{A}$, the empirical distribution of $\delta$ is shown in~\cref{fig:distribution-phase}(a). 

\medskip\noindent\emph{Stage 4.\ Noise removal.}\nopagebreak

As mentioned in Assumption~(6), we assume that $\theta_\text{A} = 0$ is ideal, and the measurement result in this case characterises the fluctuation from the remaining parts of the experiment, which we treat as noise. To obtain the real phase distributions for $\theta_\text{A}=\pi/2$ and $\pi$, we have to filter out this noise from the measured results. This is done by using a SVD filter~\cite{grassberger1993,konstantinides1997} (see \cref{sec:SVD}), with help from the characterised distribution of noise. In this processing, the SVD algorithm is first applied to the signal of the measured Gaussian noise to factorize out the singular values~(the positive square root of eigenvalues) of the noise, which contributes to small singular values~\cite{yang2003}. Then a filtering threshold is set. Consequently, the SVD algorithm is employed to the signal measured by CD1 and CD2 in the cases $\theta_\text{A}=\pi/2$ and $\pi$. The obtained singular values that are smaller than the threshold are discarded to remove the effect of the noise. The remaining singular values are reconstructed by the reverse SVD to form the filtered signal.

Then the phase distribution is calculated again, as is shown in~\cref{fig:distribution-phase}(b). We note that thanks to Assumption~(6), we can assume that the distributions in \cref{fig:distribution-phase}(b) represent the real distributions of phases that are modulated by Alice. Since the distributions are assumed to be Gaussian, which looks indeed to be the case in \cref{fig:distribution-phase}, we describe the real phase $\theta'_\text{A} =\theta_\text{A} + \delta$ by a mean value~$\overline{\theta'_\text{A}}$ and a standard deviation~$\sigma_{\theta'_\text{A}}$, listed in~\cref{tbl:parameters-phase}. Here $\overline{\theta'_\text{A}}$ represents the amount of systematic error remaining after a careful calibration process and $\sigma_{\theta'_\text{A}}$ quantifies the actual random fluctuation of the optical pulses' phase.

\begin{table}
\vspace{-0.8em} % compensates for REVTeX layout bug
\caption{Parameters of Gaussian approximation of phase distributions.}
\label{tbl:parameters-phase}
\begin{tabular}[t]{c@{\quad}c@{\quad}c}
	\hline\hline
	Nominal phase~$\theta_\text{A}$&$\overline{\theta'_\text{A}}$&$\sigma_{\theta'_\text{A}}$\\
	\hline
	0 & 0 & 0\\
	$\pi/2$ & $0.4970\pi$ & $0.0028\pi$ \\
	$\pi$ & $0.9939\pi$ & $0.0057\pi$ \\
	\hline\hline
\end{tabular}                             
\end{table}

While Clavis2 does not self-calibrate the voltages applied at the phase modulators after leaving the factory, other QKD systems may periodically recalibrate them, making $\overline{\theta'_\text{A}}$ vary during the operation. They may also drift with changing environmental conditions (such as the temperature) and the age of the hardware. However, a minor deviation of the mean from the ideal value has virtually no effect on the key rate in this security proof~\cite{mizutani2019}.

\section{Experiment of intensity characterisation}
\label{sec:experiment-intensity}

In this section, we present the methodology and experimental results of measuring the intensity fluctuation. Similar to the phase interval, the characterised intensity interval $R^k_{\rm int}$ is applied to estimate the lower bound on the single-photon detection rate among the signal states according to Eq.~(21) in Ref.~\onlinecite{mizutani2019}, which is then employed in the framework of security proof proposed in Ref.~\onlinecite{mizutani2019}. To demonstrate the method of characterising the intensity interval, we conduct a proof-of-principle experiment on another industrial-prototype BB84 QKD system that employs weak~+~vacuum decoy-state protocol~\cite{ma2005} and polarization encoding. (Since Clavis2 lacks intensity modulation and this BB84 system uses polarization encoding, we could not measure both phase and intensity fluctuation in the same system. However using two different systems should not impair the demonstration of either characterization methodology.) Here we measure $R^k_{\rm int}$ for the signal, decoy, and vacuum state.

\begin{figure}
\includegraphics{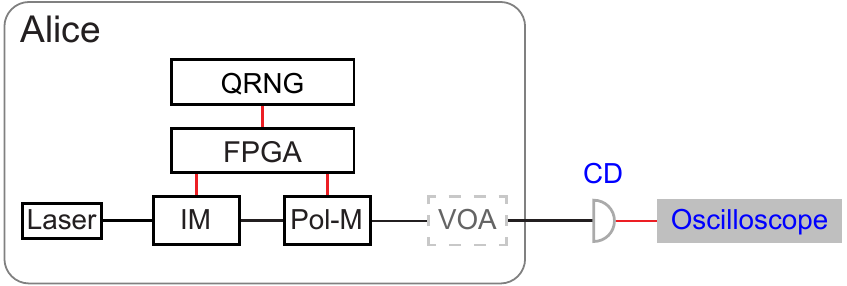} 
\caption{\label{fig:setup-intensity}Setup for intensity distribution measurement. In the device under test, Alice's laser diode generates optical pulses. They pass through an intensity modulator (IM) and polarization modulator \mbox{(Pol-M)}, which are both controlled by the FPGA with random numbers provided by QRNG. The VOA in Alice is inactivated by bypassing it. The measurement devices are classical photodetector (CD; optical-to-electrical converter) and oscilloscope.}
\end{figure}

In our test, we only employ the source unit Alice of this QKD system, whose scheme is shown in~\cref{fig:setup-intensity}. In Alice, a laser diode generates identical optical pulses at $40~\mega\hertz$ repetition rate, which are randomly modulated to different intensities by an intensity modulator~(IM). These optical pulses are then encoded into different polarization states by a polarization modulator~\mbox{(Pol-M)} and attenuated by a VOA. The randomization of the polarization modulation and intensity modulation is provided by the FPGA with random numbers generated by QRNG. We assume any of the above parts except the VOA may contribute to the intensity fluctuation of the source. The VOA, being a slow electromechanical device, should be unable to introduce fast fluctuations. In order to be able to characterise the intensity fluctuations, we set the VOA at its minimum attenuation, obtaining brighter pulses. The output of Alice is sent to an O/E converter (Picometrix \mbox{PT-40A} with $38$-$\giga\hertz$ electrical bandwidth) that connects to an oscilloscope (Agilent DSOX93304Q with $33$-$\giga\hertz$ analog bandwidth and $80$-$\giga\hertz$ sampling rate). The system is set to run the raw key exchange during our test.

The procedure of the experiment is as follows.

\medskip\noindent\emph{Stage 1.\ Noise calibration.}\nopagebreak

\begin{figure*}
\includegraphics{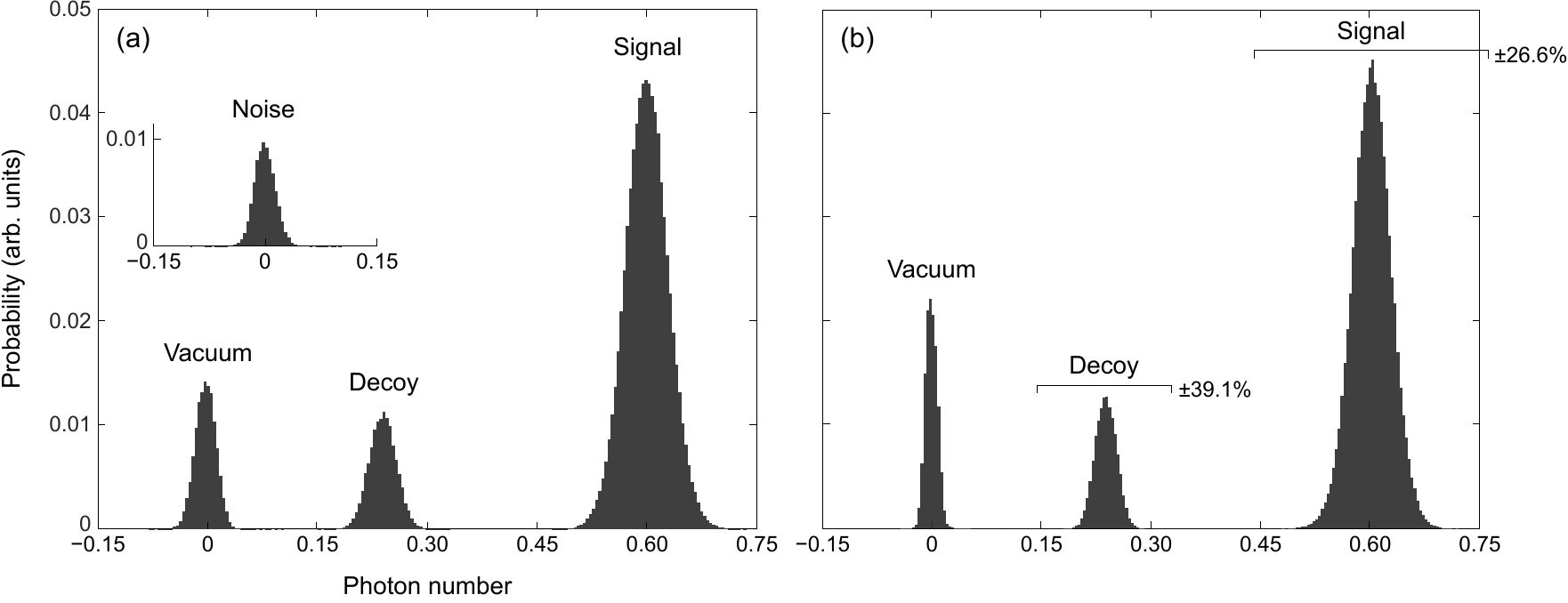}
\caption{\label{fig:distribution-intensity}Distribution of intensities for different states Alice prepares (a)~before and (b)~after noise filtering. Inset: noise distribution. Intervals $R^k_{\rm int}$ are shown as bars.}
\end{figure*}

Before conducting the intensity interval characterisation, the distribution of instrument noise shall be calibrated first. Thus, after configuring the testing setup, all the equipment, including Alice, is powered on, but no raw key exchange is started. The oscilloscope collects the distribution of instrument noise as a reference, which is shown in the inset in \cref{fig:distribution-intensity}(a). 

\medskip\noindent\emph{Stage 2.\ Measurement.}\nopagebreak

\begin{figure}
\includegraphics{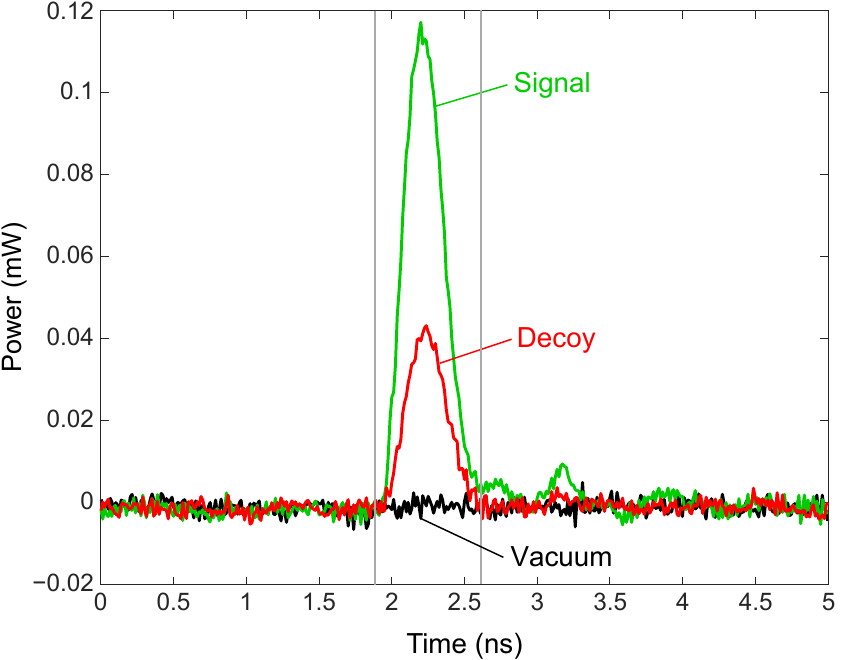}
\caption{\label{fig:waveform-intensity}Typical waveforms of the three intensity states measured in the experiment. The energy of each pulse is calculated by integrating the power between the grey lines.}
\end{figure}

In order to measure the intensity fluctuations of the signal, decoy, and vacuum state, Alice continuously runs the raw key exchange to generate optical pulses with random intensities (with signal\,:\,decoy\,:\,vacuum probability ratio of $6\!:\!1\!:\!1$), until $2\times10^5$ pulses in total are collected by the oscilloscope. In this way, the three intensity levels are recorded in one run. The typical waveforms measured in the experiment are shown in~\cref{fig:waveform-intensity}.

\medskip\noindent\emph{Stage 3.\ Intensity calculation.}\nopagebreak

After the measurement, we process the recorded oscillograms to calculate the single-photon-level intensities. The photon number of each optical pulse is calculated as
\begin{equation}
\label{int_eq}
\mu = \frac{\left(S_\text{e} / G_\text{o-e}\right) A}{h c / \lambda},
\end{equation}
where the measured area under the voltage oscillogram of each pulse $S_\text{e}$ is obtained by integrating the signal between the grey lines in~\cref{fig:waveform-intensity}, the conversion gain of the O/E converter $G_\text{o-e} = 703~\volt/\watt$ as calibrated by its manufacturer, and normal transmission of the VOA including its insertion loss in the QKD system during the raw key exchange $A = 1.876 \times 10^{-6}$ ($-57.3~\deci\bel$). The denominator is the energy of a single photon at the laser wavelength $\lambda = 1550.12~\nano\meter$, with $h$ being the Planck constant and $c$ the speed of light. The calculated pulse intensities are then binned in histograms to produce distributions shown in~\cref{fig:distribution-intensity}(a). 

\medskip\noindent\emph{Stage 4.\ Noise removal.}\nopagebreak

To remove instrument noise from the measured results, we use a SVD filter (see \cref{sec:SVD}) with an input of the instrument noise distribution collected at \emph{Stage~1}. The SVD processing is identical to that described in~\emph{Stage~4} of the phase interval measurement. The processing is applied to the measured electrical signals of all three intensity levels. The filtered electrical signals are then used to calculate the photon numbers again according to \cref{int_eq}, which are binned into histograms again. The result is shown in~\cref{fig:distribution-intensity}(b). Similarly to the phase distributions, the distributions of intensities are also nearly Gaussian, with their parameters listed in~\cref{tbl:parameters-intensity}. It is notable that, in theory, the vacuum state is zero. However, in practice, measurement is always affected by noise, so we cannot get the perfect zero for the vacuum state but obtain, in this particular instance, a small negative value.   

\begin{table}
\vspace{-0.8em} % compensates for REVTeX layout bug
\caption{Parameters of Gaussian approximation of intensity distributions.}
\label{tbl:parameters-intensity}
\begin{tabular}[t]{c@{\quad}c@{\quad}c}
	\hline\hline
	State & $\bar\mu$ & $\sigma_\mu$ \\
	\hline
	Vacuum	& $-0.78\times10^{-3}$	& 0.0083 \\
	Decoy		& 0.236									& 0.0149 \\
	Signal	& 0.602									& 0.0258 \\
	\hline\hline
\end{tabular}
\end{table}

The extracted intensity distributions are much wider relative to their mean values than the phase distributions. We attribute this to stochastic dynamic processes in the gain-switched laser that generate energy noise (and timing jitter) of short pulses produced by it. As will be shown shortly, this leads to zero secure key rate with the available proof. This indicates that this simple gain-switched laser source may be unsuitable for secure QKD. Improvements to the source that reduce the laser's timing jitter \cite{taofiq2021} might also reduce its energy noise and should be tested in future work. Moreover, the random modulation of intensity, compared to the fixed modulation of phase, may also introduce extra fluctuation, which also shall be investigated in future work.

\section{Simulation of secret key rate}
\label{sec:simulation}

We employ the security proof and simulation technique proposed in Ref.~\onlinecite{mizutani2019} to calculate the secret key rate. In order to thoroughly show the effect of the phase and intensity fluctuation on the secret key rate, we consider three cases. We first calculate the key rate with phase fluctuation only (assuming there is no intensity fluctuation), then with intensity fluctuation only, then with both phase and intensity fluctuation. For all the following simulations we assume that Bob uses single-photon detectors with $80\%$ photon detection efficiency and dark count probability of $10^{-9}$ per optical pulse, such as superconducting-nanowire detectors from Scontel~\cite{Scontel}. The security parameter $\varepsilon_s$ is set to be $10^{-10}$, and the efficiency of the error correcting codes $f_e$ is assumed to be $1.05$. We assume fiber loss of $0.20~\deci\bel\per\kilo\meter$. 

\subsection{Phase fluctuation only}

\begin{figure}
\includegraphics{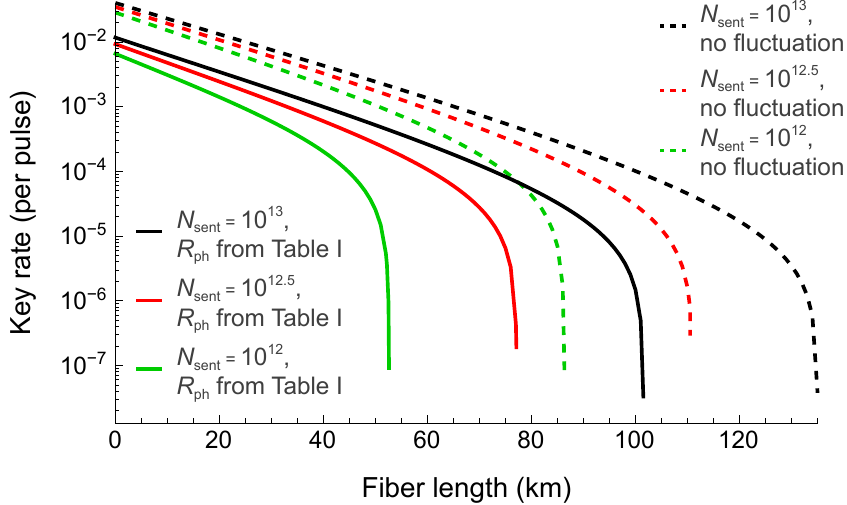}
\caption{\label{fig:keyrate-phase}Simulated key rate when only phase fluctuation is considered. The dashed lines are key rates with the measured mean value of phase but without any fluctuation. The solid lines show key rates with the phase fluctuation obtained from experiment in \cref{tbl:parameters-phase}. The number of pulses sent by Alice is $10^{13}$ for black lines, $10^{12.5}$ for red (dark grey) lines, and $10^{12}$ for green (light grey) lines.}
\end{figure}

We first simulate the scenario when the intensity modulation is perfect and only the phase fluctuation exists. We apply the experimental values of phase fluctuation from \cref{tbl:parameters-phase}. The phase falling outside the fluctuation interval $\overline{\theta'_\text{A}} \pm 6.2 \sigma_{\theta'_\text{A}}$ either for $\theta_A=0$ or $ \pi/2$ is regarded as an error, and the probability of having such error $p_\text{fail} = 1.13 \times 10^{-9}$. Signal and decoy intensities are optimised at each fiber length. The simulation results are shown in \cref{fig:keyrate-phase}. The dashed (solid) lines show the secret key rates without (with) phase fluctuation, for three different total numbers of pulses sent by Alice $N_\text{sent}$. 

The presence of the phase fluctuation reduces the secure key rate at a short distance by about a factor of four and the maximum transmission distance by about $35~\kilo\meter$. Notably, the distance can still be longer than $100~\kilo\meter$ for $10^{13}$ pulses sent. Although in Clavis2 a key distribution session of the latter size would take about two months, a higher-speed $1~\giga\hertz$ system could complete it under $3~\hour$.

To see how much phase noise the system can tolerate, we have also run the simulation with an artificially increased fluctuation, multiplying the experimentally obtained $\sigma_{\theta'_\text{A}}$ by a factor. The system keeps producing secret key until $\sim 2 \sigma_{\theta'_\text{A}}$.

\subsection{Intensity fluctuation only}

Next, we calculate the key rates with intensity fluctuation but no phase fluctuation. Unfortunately, the simulation cannot directly take the measured mean photon numbers and standard deviations shown in~\cref{tbl:parameters-intensity} to obtain a positive key rate. This is because the measured mean photon numbers of the decoy and signal are not optimised for the simulation at each fiber length. Moreover, the measured intensity fluctuation of the gain-switched laser diode source is too large to generate a key under this proof. Therefore, in order to provide a reference for an acceptable level of intensity fluctuation, we analyse the key rates under two scenarios: (a)~the mean photon numbers are fixed and taken from the experimental results but the fluctuation interval is set to be $\pm 0.5 \%$ or $\pm 1 \%$, and (b)~the mean photon numbers are optimised for each fiber length and fluctuation is also set to be $\pm 1 \%$. In both scenarios, we set the probability that the actual intensity value is outside the fluctuation interval at $1.41 \times 10^{-9}$. The vacuum state is assumed to have $\bar\mu = 10^{-3}$ and zero fluctuation. The phase is set to have zero fluctuation but the measured mean values $\overline{\theta'_\text{A}}$ (\cref{tbl:parameters-phase}); the latter values have virtually no effect on the key rate in this security proof~\cite{mizutani2019}.

\begin{figure}
\includegraphics{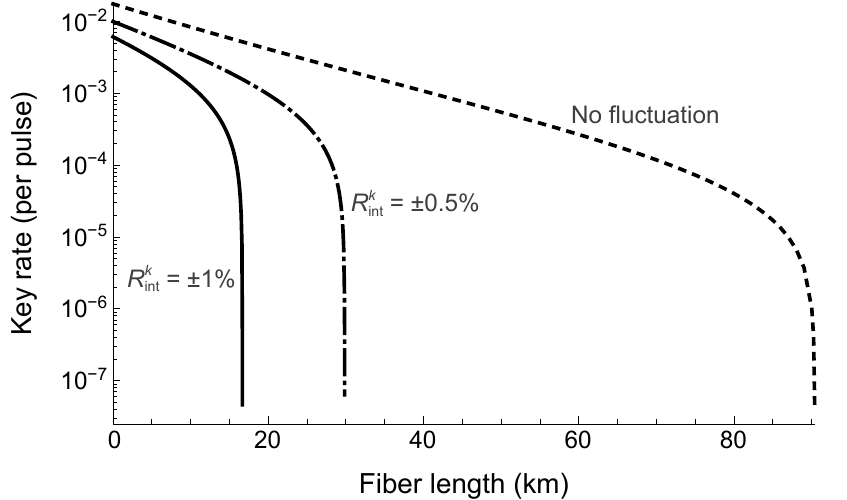}
\caption{\label{fig:keyrate-intensity-fixed}Simulated key rates with only intensity fluctuation and fixed mean photon numbers obtained from the experiment. The dashed line is the key rate with no fluctuation at all. Then dash-dotted (solid) line is the key rate with $\pm 0.5\%$ ($\pm 1\%$) fluctuation of intensity modulation. The number of pulses sent by Alice is $10^{13}$.}
\end{figure}

\emph{Scenario~(a): Fixed mean photon numbers with fixed fluctuation.} The mean photon numbers for the decoy and signal are taken from~\cref{tbl:parameters-intensity}, and $N_\text{sent} = 10^{13}$. The resulting key rates with different amounts of intensity fluctuation are shown in~\cref{fig:keyrate-intensity-fixed}. The fluctuation is set to be $\pm 0.5\%$ or $\pm 1\%$ for both the signal and decoy state. While without the imperfection the distance reaches $90~\kilo\meter$, a moderate amount of fluctuation of just $\pm 0.5\%$ ($\pm 1\%$) reduces it to $30~\kilo\meter$ ($17~\kilo\meter$). This high sensitivity to the fluctuation is attributed to our using the fixed mean photon numbers of the states. Note that the fluctuation ranges simulated are $\sim 25$--$80$ times smaller than those obtained from the experiment in~\cref{sec:experiment-intensity}: the $\pm 6.2\sigma_{\mu}$ range is $\pm 39.1\%$ for the decoy state and $\pm 26.6\%$ for the signal state. The source intensity noise should thus be drastically reduced to accommodate the available security proof.

\begin{figure}
\includegraphics{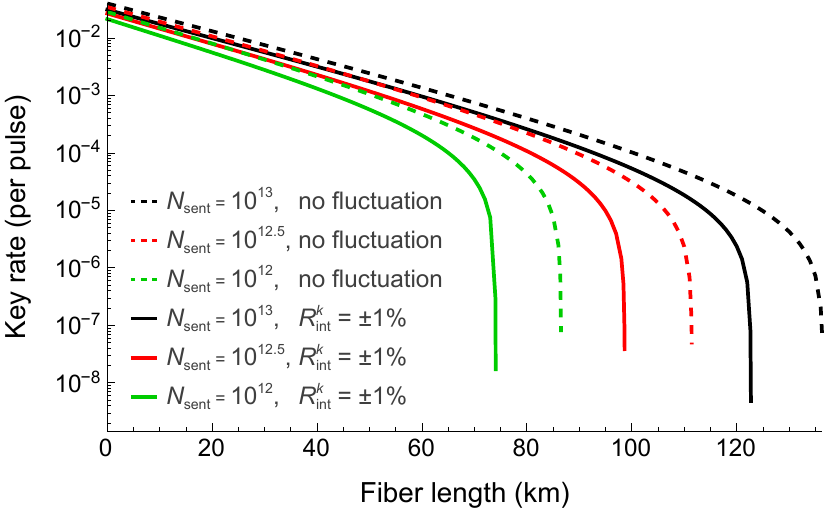}
\caption{\label{fig:keyrate-intensity-optimised}Simulated key rates with only intensity fluctuation with optimised mean photon numbers at each distance. The dashed lines are the secret key rates without any fluctuation, while the solid lines are those with $\pm 1\%$ fluctuation of intensity modulation. The number of pulses sent by Alice is $10^{13}$ for black lines, $10^{12.5}$ for red (dark grey) lines, and $10^{12}$ for green (light grey) lines.}
\end{figure}

\emph{Scenario~(b): Optimised mean photon numbers with fixed fluctuation.} In this simulation, the mean photon number of the decoy state is optimised between $0.001$ and $0.15$, and that of the signal state is optimised between $0.1$ and $0.8$. The results are shown in \cref{fig:keyrate-intensity-optimised}. Key rates for three different $N_\text{sent}$ are plotted, with all the other parameters being identical. 
 
This simulation shows that optimising the mean photon numbers of the decoy and signal allows to tolerate the intensity fluctuation much better. The key rate at a short distance is not significantly reduced and the maximum transmission distance only decreases by about $13~\kilo\meter$. For the same fluctuation of $\pm 1\%$, the maximum distance with optimised photon numbers reaches $123~\kilo\meter$ versus $17~\kilo\meter$ with the fixed photon numbers. This is a significant advantage. However this means the experimental characterisation should in the future be extended to measure fluctuation in a range of mean photon numbers, which makes it more complex.

\subsection{Phase and intensity fluctuation}

\begin{figure}
\includegraphics{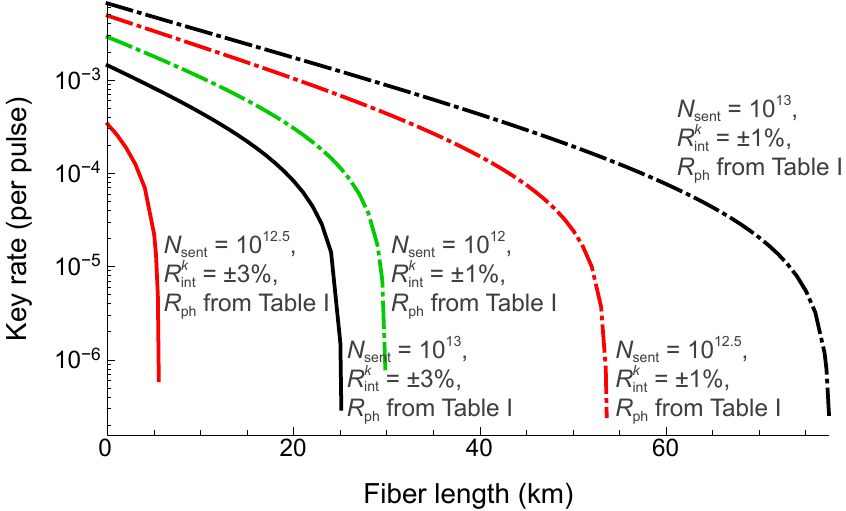}
\caption{\label{fig:keyrate-intensity-and-phase}Simulated key rates when both phase and intensity fluctuations are considered. The phase fluctuations are obtained from the experiment (\cref{tbl:parameters-phase}) and the mean photon numbers of the decoy and signal state are optimised at each fiber length. The dash-dotted (solid) lines are the key rate with $\pm 1\%$ ($\pm 3\%$) intensity fluctuation. The number of pulses sent by Alice is $10^{13}$ for black lines, $10^{12.5}$ for red (dark grey) lines, and $10^{12}$ for green (light grey) lines.}
\end{figure}

Finally, we consider fluctuations in both the intensity and phase. The overall error probability is set to be $2.54 \times 10^{-9}$, which is the probability that the phase is outside the fluctuation interval of $\pm 6.2 \sigma_{\theta'_\text{A}}$ or the actual intensity value is outside the fluctuation interval of $\pm 1\%$ or $\pm 3\%$. Other relevant assumptions are taken from the phase-only and intensity-only fluctuation simulations. The results are shown in \cref{fig:keyrate-intensity-and-phase}.

Combining the phase and intensity fluctuation introduces some drop in the key rate and maximum distance. For example, with only intensity fluctuation of $\pm 1\%$ interval, the maximum distance is $123~\kilo\meter$ for $N_\text{sent} = 10^{13}$ (\cref{fig:keyrate-intensity-optimised}), however when the phase fluctuation is added it drops to $78~\kilo\meter$ (\cref{fig:keyrate-intensity-and-phase}). When the intensity interval is widened to $\pm 3\%$, the key rate and maximum distance decay rapidly, and no key is produced for the lowest $N_\text{sent} = 10^{12}$. This shows that controlling the intensity fluctuation in the QKD hardware is crucial, at least with the security proof currently available \cite{mizutani2019}.

We stress that our security proof \cite{mizutani2019} assumes that when an individual state is outside the set fluctuation range, Eve can get perfect information about the raw key, which is the worst scenario in theory but is not a specific known attack. We've admitted this for the simplicity of analysis, and this assumption may be one of the major causes of the severe performance degradation. In the future, we plan to consider how to remove this assumption for better performance.

\section{Conclusion}
\label{sec:conclusion}

We have proposed and experimentally demonstrated methodology for characterising source fluctuation in phase and intensity in QKD. We have then applied our characterisation results to the security proof of the three-state, loss-tolerant protocol~\cite{mizutani2019}. The fluctuations lead to a significant reduction in the secure key rate and maximum transmission distance in fiber. In fact, the intensity fluctuation we measured on the gain-switched semiconductor laser source is so large that the proof predicts no key. There is room for improvement in the source hardware, especially to reduce its intensity fluctuation. An alternative might be using a seeded pulsed laser source \cite{taofiq2021} or using a continuous-wave laser and produce phase-randomised pulses with intensity and phase modulators. Likewise the security proof and details of the QKD protocol might be improved to give a higher key rate at the same fluctuation. This may be helped by knowing the distribution of fluctuation (such as the Gaussian distribution we measured) \cite{sixto2022} or by combining with the idea of twisting~\cite{bourassa2020}. Finally, the development of security proofs for other QKD protocols that incorporate source fluctuation is desirable \cite{wang2022,gu2022}. Although the security proof we use~\cite{mizutani2019} cannot handle intersymbol interference~\cite{yoshino2018,pereira2020,mizutani2021,zapatero2021,navarrete2021}, our experimental characterisation techniques can be adapted to also measure the latter. This characterisation methodology is a necessary element in the upcoming formal security standards and certification of QKD.

\acknowledgments
We thank Rikizo Ikuta, Yongping Song, Weixu Shi, Dongyang Wang, Jing Ren, and Qingquan Peng for helpful discussions. We thank ID~Quantique for cooperation and a loan of equipment. This work was funded by Canada Foundation for Innovation, MRIS of Ontario, the National Natural Science Foundation of China (grants 61901483 and 62061136011), the National Key Research and Development Program of China (grant 2019QY0702), the Research Fund Program of State Key Laboratory of High Performance Computing (grant 202001-02), and the Ministry of Education and Science of Russia (program NTI center for quantum communications). H.-K.L.\ was supported by NSERC, MITACS, ORF, and the University of Hong Kong start-up grant. V.M.\ was supported by the Russian Science Foundation (grant 21-42-00040). K.T.\ was supported by JSPS KAKENHI grant JP18H05237 and JST CREST grant JPMJCR 1671.

This article was disclosed to our affected industry partners prior to its publication.

{\em Author contributions:} A.H.\ performed the experiments. A.M.\ and K.T.\ performed the simulations. H.-K.L.,\ V.M., and K.T. supervised the study. A.H.\ wrote the article with help from all authors.

\appendix

\setcounter{figure}{0}
\numberwithin{figure}{section}
\numberwithin{equation}{section}

\section{Protocol framework}
\label{sec:protocol}

We briefly describe the protocol~\cite{mizutani2019} that the security proof with modulation fluctuation is based on. Here we consider the loss-tolerant protocol~\cite{tamaki2014} with decoy-state method, which uses three encoding states, $\{0_Z, 1_Z, 0_X\}$, with asymmetric choice between two bases. Additionally, we follow the framework of security proof in Ref.~\onlinecite{mizutani2019}, and the secret key is extracted from the events where Alice and Bob both select Z basis. The detection events where Alice and Bob have chosen different bases are used for estimating the phase error rate, i.e., the amount of information leaked to the eavesdropper.

Before describing the protocol~\cite{mizutani2019}, we summarise the assumptions on the light source. How our experiment allows to satisfy these assumptions and extract the values for the relevant parameters is explained in main text. The following labels (A-1)--(A-7) correspond to the ones in Ref.~\onlinecite{mizutani2019}. Note that the main contribution of our experiment is to measure the phase intervals $R^c_{\rm ph}$ and intensity intervals $R^k_{\rm int}$ in (A-6).
\\
(A-1) Assumption on the emitted state:
We assume that each emitted state is a perfectly-phase-randomised and single-mode coherent state. We just assume that our experiments with both QKD systems satisfy this assumption.
\\
(A-2) Correlation assumption:
We assume that each emitted signal is allowed to be correlated in a setting-choice-independent correlation (SCIC) manner~\footnote{Especially in high-speed QKD systems, each emitted signal could be correlated in a setting-choice-dependent correlation (SCDC) manner, which means that setting-choice information propagates to subsequent pulses.  Some security proofs accommodate this type of correlation; intensity correlations are accommodated in Refs.~\onlinecite{yoshino2018,zapatero2021}, and pulse correlations in terms of Alice’s bit choice information are taken into account in Refs.~\onlinecite{pereira2020,mizutani2021,navarrete2021}.}. Note that SCIC means that there exists a parameter $g^i$ in the source device that determines the $i$-th emitted state, and these $\{g^i\}$ are allowed to be correlated with each other. For example, $g^i$ is a temperature of the source device of the $i$-th pulse emission.
\\
(A-3) Random choice assumption:
We assume that each intensity choice $k \in \mathcal{K}:=\{k_1,k_2,k_3\}$ and each phase choice $c \in \mathcal{C}:=\{0_Z,1_Z,0_X\}$ are independent random variables with fixed chosen probabilities.
\\
(A-4) Independence assumption of the phase and the intensity:
We assume that the $i$-th phase and intensity are independent from one another, which we just suppose in our experiment.
\\
(A-5) Unique determination assumption of the phase and the intensity:
Given the intensity and phase choices and the parameter $g^i$, the $i$-th phase and intensity are uniquely determined, which we also just assume in our experiment.
\\
(A-6) Assumption on the intervals for the phase and intensity:
For each phase choice $c$, Alice knows the phase interval $R^c_{\rm ph}$ and for each intensity choice $k$, she knows the intensity interval $R^k_{\rm int}$. The contribution of this paper is to characterise $R^c_{\rm ph}$ and $R^k_{\rm int}$ in practical QKD systems.
\\
(A-7) Assumption on the number of tagged signals:
Among all the emitted pulses, the number of tagged signals $n_{\rm tag}$ are assumed to be upper-bounded by the probability $p_{\rm fail}$. Again, the tagged signals are the emitted pulses whose intensity or phase is outside the intervals. This is just a theoretical assumption. We can thus freely choose the number of tagged signals $n_{\rm tag}$ and the probability $p_{\rm fail}$ of their falling outside the intervals $R^c_{\rm ph}$ and $R^k_{\rm int}$, such that we obtain the maximum key rate.
\\

Then, the loss-tolerant protocol in Ref.~\onlinecite{mizutani2019} operates as follows.

\medskip\noindent {\it Step 0: Device characterisation and parameter determination.} Alice first characterises her phase fluctuation interval, $R_\text{ph}^{c}$, and intensity fluctuation interval, $R_{\rm int}^k$, where $c \in \mathcal{C} = \{0_Z, 1_Z, 0_X\}$ for different encoding states and $k \in \mathcal{K} = \{k_1, k_2,k_3\}$ for different intensities. Furthermore, Alice and Bob shall decide the security parameter $\varepsilon_s$, and the number of pulses sent $N_\text{sent}$.

\medskip\noindent {\it Step 1: State preparation and transmission.} Alice randomly chooses an intensity setting 
$k \in \mathcal{K}$ and a basis with probability $P \in \{P_Z, P_X \}$. If Z basis is selected, she prepares bit $0_Z$ or bit $1_Z$ with equal probability. Otherwise, she chooses bit $0_X$. The prepared state is sent to Bob via a quantum channel.

\medskip\noindent {\it Step 2: Detection.} Bob measures the received states by randomly choosing a basis from {Z, X} with probability $P \in \{P_Z, P_X \}$. The detection outcome is $b \in \{0, 1, \bot, \emptyset \}$, where $\bot$ and $\emptyset$ represent a double-click event and no-click event respectively. In the double-click event, a bit value is randomly assigned.

Alice and Bob repeat Step~1 and Step~2 until $N_\text{sent}$ pulses are sent to Bob.

\medskip\noindent {\it Step 3: Sifting.} Bob declares the detection events and his basis choices for these events over an authenticated public channel. Alice checks her basis choices during these events and announces to Bob. They keep the events where both of them have selected Z basis as a sifted key. 

\medskip\noindent {\it Step 4: Parameter estimation.} Alice uses a subset of the sifted key and the intensity interval $R_{\rm int}^k$ to calculate a lower bound on the number of single-photon events $S_{Z,1}^L$, by Eq.~(21) in Ref.~\onlinecite{mizutani2019}. Moreover, she calculates an upper bound on the number of phase errors $N_{ph,Z,1}^U$ with help of phase interval $R_\text{ph}^{c}$, according to Eq.~(24) in Ref.~\onlinecite{mizutani2019}. Thus, an upper bound on the phase error rate is given by $e_{ph,Z,1}^U = N_{ph,Z,1}^U/S_{Z,1}^L$.

\medskip\noindent {\it Step 5: Error correction.} Through the public channel, Bob corrects his sifted key to be identical to that of Alice.

\medskip\noindent {\it Step 6: Privacy amplification.} Alice and Bob conduct privacy amplification to shorten the key length by using the results from Step~4 and Step~5 to obtain the final secret key with length $l$, which satisfies~\cite{mizutani2019}
\begin{equation}
\label{key rate}
l \leq S_{Z,1}^L \left[1-h(e_{ph,Z,1}^U)\right] - \log_2\frac{2}{\varepsilon_\text{PA}} -\lambda_{\rm EC},
\end{equation}
where $h(x)$ is the binary entropy function, $\varepsilon_\text{PA} > 0$ is the parameter related to the success probability of privacy amplification, and $\lambda_\text{EC}$ is the cost of error correction.

\section{Denoising by singular value decomposition}
\label{sec:SVD}

The instrument noise produced by our characterisation setups is assumed to be additive and independently-and-identically-distributed (i.i.d.)\ Gaussian, or the white noise. We would like to filter out this uncorrelated instrument noise from the measured phase and intensity fluctuations produced by the QKD systems under test. If these phase and intensity fluctuations also had Gaussian distribution, their denoising would be very simple. The denoised distribution would be Gaussian with the same mean as the measured distribution and the standard deviation $\sigma = \sqrt{\sigma_\text{measured}^2-\sigma_\text{noise}^2}$, where $\sigma_\text{measured}$ is the standard deviation of the measured fluctuation-with-instrument-noise distribution and $\sigma_\text{noise}$ is the standard deviation of the measured instrument-noise-only distribution.

However if the phase and intensity fluctuations might not follow the Gaussian distribution, we need to adopt a more general denoising technique. In this case, filtering has to be applied to individual measured waveforms (shown in~\cref{fig:waveform-phase-pi-2,fig:waveform-intensity}). In our study, we apply a filtering process based on singular value decomposition (SVD)~\cite{grassberger1993,konstantinides1997,jha2011}. The measured waveforms are put into rows in an $m \times n$ matrix $M$, which is then factorised as
\begin{equation}
\label{eq:SVD}
M = USV\tran,
\end{equation}
where $U$ is an $m \times m$ unitary matrix of orthonormal eigenvectors of $MM\tran$, $S$ is an $m \times n$ diagonal matrix of singular values that are the square roots of the eigenvalues of $M\tran M$ and are arranged in descending order, and $V\tran$ is an $n \times n$ unitary matrix containing the orthonormal eigenvectors of $M\tran M$. After this SVD processing, the data contained in the matrix $M$ is represented by the singular values in matrix $S$.

The singular values and corresponding singular vectors contain complete information about matrix $M$. The characteristics of the waveforms are mainly described by the first few singular values in matrix $S$. Meanwhile, the small singular values of the additive instrument noise are assumed to be spread over its $m$ dimension~\cite{yang2003}. Therefore, theoretically the SVD method could divide the data space of the measured waveforms into true signal and noise subspaces by distinguishing and separating the singular values. Specifically, for the instrument noise that is i.i.d.\ Gaussian, a threshold could be applied to distinguish between small singular values of the noise and large ones of the signal. This threshold is determined taking into account the prior knowledge about the strength of the instrument noise~\cite{workalemahu2008}. We then truncate (i.e.,\ zero) the small singular values that fall below the threshold. Practice shows that the SVD method can be effective at filtering out the noise. The specific algorithm we use is the following.

\medskip\noindent
{\it Step~0.~Construct the matrix of waveforms.} After obtaining the measured waveforms of optical pulses detected by the O/E converter as oscillograms (shown in~\cref{fig:waveform-phase-pi-2,fig:waveform-intensity}), each waveform period constitutes a row of the matrix $M$. Let's consider the intensity fluctuation as an example. Since the repetition frequency of the measured QKD system is $40~\mega\hertz$ and the sampling rate of the oscilloscope is $80~\giga\hertz$, each period contains 2000 sampling points, which populate a row of the matrix $M$. To limit the computing time of SVD, our data is split into multiple matrices containing each $100$ periods of oscillogram's data, so that the size of each matrix $M$ is $100 \times 2000$. In principle the waveforms could be processed in fewer matrices of a larger size, however this would take more computing time.

\medskip\noindent
{\it Step~1.~Obtain singular values.} The singular value decomposition is applied to each matrix $M$, obtaining the matrix $S$ with singular values on the diagonal. For example, in the case of intensity fluctuation, the SVD is first applied to the waveforms of instrument noise, which allows us to know all the singular values of the noise. The instrument noise waveforms are processed in blocks of $100$, then their singular values are averaged over all the blocks to obtain a reference set. Then the waveforms of optical pulses for the vacuum, decoy, and signal state are processed by SVD to obtain their singular values as well (individually for each block, without averaging). \Cref{fig:SVD} shows the first $50$ singular values for the measured waveforms of the three latter states in one of the blocks. It can be seen that only a few singular values are dominating elements with information about the intensity, and the remaining singular values are relatively smaller due to the i.i.d.\ instrument noise.

\medskip\noindent
{\it Step~2.~Determine the truncation threshold.} A truncation threshold $\epsilon$ is manually determined either to be the singular value that suddenly becomes smaller than the previous one or the singular value that is close to the maximum one of the instrument noise. The former approach is used when a signal-to-noise ratio (SNR) is high, such as for the waveforms of optical pulses for the decoy and signal state. The latter approach is used when the SNR is relatively low, such as for the vacuum state. In our measurement of intensity fluctuation, the maximum singular value of the instrument noise in the reference set is $0.4691$, which we regard as the truncation threshold of the vacuum state. Thus, for the signal (decoy, vacuum) state, $\epsilon$ is set to be the value of the seventh (fifth, twelfth) element, as illustrated in~\cref{fig:SVD}.

\medskip\noindent
{\it Step~3.~Truncate and reconstruct the filtered waveform.} The singular values that are equal to or smaller than $\epsilon$ are zeroed, obtaining matrix $S'$. The filtered waveform is then reconstructed as $M' = US'V\tran$. After that, the energy of each optical pulse is calculated by integrating the filtered waveform, whose data is each row of matrix $M'$. The fluctuation distributions after filtering are then calculated according to {\it Stage~3} presented in main text for the cases of both phase fluctuation and intensity fluctuation.

\renewcommand{\thefigure}{12}
\begin{figure}
\includegraphics{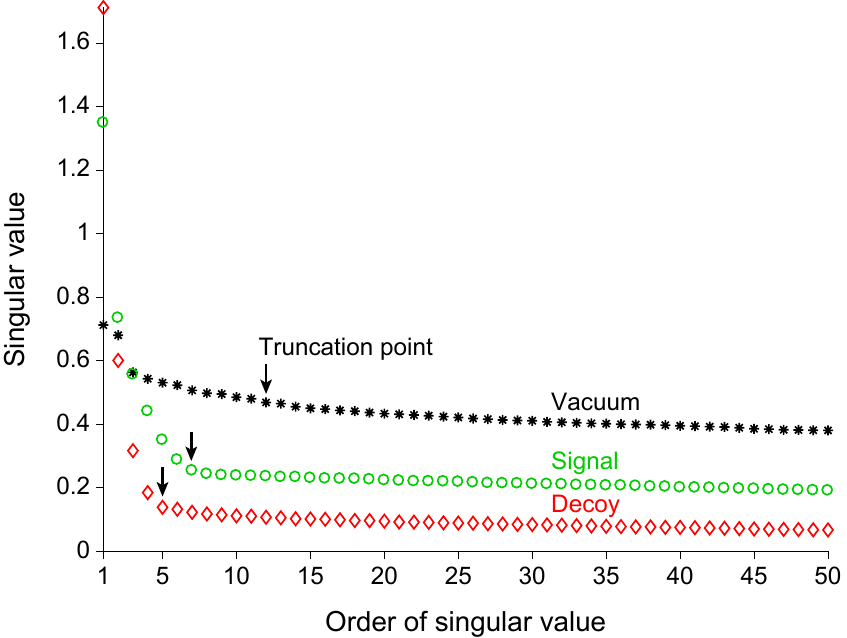}
\caption{\label{fig:SVD}The first $50$ singular values obtained by SVD for the measured waveforms of optical pulses.}
\end{figure}

We remark that for the measurements presented in this study in \cref{fig:distribution-phase,fig:distribution-intensity}, the distributions filtered by SVD happen to be close to Gaussian. The simple Gaussian denoising technique described in the beginning of this Appendix gives a result that is close to that of SVD. However the SVD can also handle non-Gaussian experimentally measured distributions (which we have observed in some of our experiments), thus it remains our chosen denoising method.

\def\bibsection{\medskip\begin{center}\rule{0.5\columnwidth}{.8pt}\end{center}\medskip}% Redefines bibliography separator to single-column. This reduces chances of float placement bugs in the last page.
\bibliography{library}

\end{document}